# UV surface brightness of galaxies

# from the local Universe to z ~ 5

Eric J. Lerner
*Lawrenceville Plasma Physics, Inc., USA*
eric@LPphysics.com

Renato Falomo
*INAF – Osservatorio Astronomico di Padova, Italy*
renato.falomo@oapd.inaf.it

and

Riccardo Scarpa
*Instituto de Astrofisica de Canarias, Spain*
riccardo.scarpa@gtc.iac.es

**ABSTRACT**

The Tolman test for surface brightness dimming was originally proposed as a test for the expansion of the Universe. The test, which is independent of the details of the assumed cosmology, is based on comparisons of the surface brightness (SB) of identical objects at different cosmological distances. Claims have been made that the Tolman test provides compelling evidence against a static model for the Universe. In this paper we reconsider this subject by adopting a static Euclidean Universe with a linear Hubble relation at all z (which is not the standard Einstein − de Sitter model), resulting in a relation between flux and luminosity that is virtually indistinguishable from the one used for ΛCDM models. Based on the analysis of the UV surface brightness of luminous disk galaxies from HUDF and GALEX datasets, reaching from the local Universe to $z \sim 5$, we show that the surface brightness remains constant as expected in a SEU.

A re-analysis of previously-published data used for the Tolman test at lower redshift, when treated within the same framework, confirms the results of the present analysis by extending our claim to elliptical galaxies. We conclude that available observations of galactic SB are consistent with a static Euclidean model of the Universe.

We do not claim that the consistency of the adopted model with SB data is sufficient by itself to confirm what would be a radical transformation in our understanding of the cosmos. However, we believe this result is more than sufficient reason to examine further this combination of hypotheses.

*Subject headings:* Cosmology: general, Tolman test

# 1.    Introduction

As Tolman [1,2] demonstrated, the dependence of the bolometric surface brightness (SB) of  identical objects as a function of redshift z is independent of the specific parameter of the adopted cosmology, e.g., Hubble constant, dark matter $\Omega_M$ and dark energy $\Omega_\Lambda$ content of the Universe. For this reason the comparison of the surface brightness of similar objects at different distance was seen as a powerful tool to test for the expansion of the Universe. In fact, in any expanding cosmology, the SB is expected to decrease very rapidly, being proportional to $(1+z)^{-4}$, where z is the redshift and where SB is measured in the bolometric units  (VEGA-magnitudes/arcsec$^{-2}$ or erg sec$^{-1}$cm$^{-2}$arcsec$^{-2}$). One factor of $(1+z)$ is due to time-dilation (decrease in photons per unit time), one factor is from the decrease in energy carried by photons, and the other two factors are due to the object being closer to us by a factor of $(1+z)$ at the time the light was emitted and thus having a larger apparent angular size. (If AB magnitudes or flux densities are used, the dimming is by a factor of $(1 + z)^3$, while for space telescope magnitudes or flux per wavelength units, the dimming is by a factor of $(z + 1)^5$).By contrast, in a static (non expanding) Universe, where the redshift is due to some physical process other than expansion (e.g., light-aging), the SB is expected to dim only by a factor $(1 + z)$, or be strictly constant when AB magnitudes are used.

In the last few decades the use of modern ground-based and space-based facilities have provided a huge amount of high quality data for the high-z Universe. The picture emerging from these data indicates that galaxies evolve over cosmic time. The combination of cosmological effects with the evolution of structural properties of galaxies makes the Tolman test more complicated to implement because of the difficulty in disentangling two types of effects (cosmology and intrinsic evolution). In spite of this complexity, various authors have attempted to perform the Tolman test [3,4,5], most reaching the conclusion that the Tolman test ruled out the static Universe model with high confidence.

In this paper we present a new implementation of the Tolman test based on  a comparison of the UV surface brightness of a large sample of disk galaxies from the local Universe to $z \sim 5$ as well as a critical re-analysis of previously-published data. Preliminary reports of this work were presented by Lerner [6,7] . Consistent with those preliminary reports and contrary to earlier conclusions by other authors, we here show that the surface brightness of these galaxies remains constant over the entire redshift range explored. Based on these observations, it is therefore not true that a static Euclidean Universe can be ruled out by the Tolman test.

## 2.    The adopted cosmology.

Since the SB of galaxies is strongly correlated with the intrinsic luminosity, for a correct implementation of the Tolman test it is necessary to select samples of galaxies at different redshifts from populations that have on average the same intrinsic luminosity. To do this, one is forced to adopt a relation between z and distance d in order to convert apparent magnitudes to absolute magnitudes. In this paper we are testing a static cosmology where space is assumed Euclidean and the redshift is due to some physical process other than expansion. For this study, we adopt the simple hypothesis that the relationship d= cz/H$_0$ ,

well-assessed in the local Universe, holds for all *z*. It should be noted that this cosmological model is not the Einstein-De Sitter static Universe often used in literature.

The choice of a linear relation is motivated by the fact that the flux-luminosity relation derived from this assumption is remarkably similar numerically to the one found in the concordance cosmology, the distance modulus being virtually the same in both cosmologies for all relevant redshifts . This is shown in Fig. 1 where the two relations are compared to each other and, in Fig. 2, to supernovae type Ia data. Up to redshift 7, the apparent magnitude predicted by the simple linear Hubble relation in a Static Euclidean Universe (SEU) is within 0.3 magnitude of the concordance cosmology prediction with $\Omega_M$= 0.26 and $\Omega_\Lambda$= 0.74. The fit to the actual supernovae data is statistically indistinguishable between the two formulae.

In this particular framework the bolometric luminosity L and the flux F from a source are related by the relation $F = L/[4\pi d^2(1+z)]$, where the factor (1+z) takes into account energy losses due to the redshift. When using flux per unit frequency, that is AB magnitudes, this relation further simplifies to $F = L/(4\pi d^2)$. Therefore the absolute magnitude M can be derived from the apparent magnitude m (in the AB system) using the relation: $M - m = 5 - 5Log(cz/H_0)$.

Under the assumption of a static Universe the true size R and the apparent size r of an object are linked by the standard relation $r = R/d$ , where d is the distance and r is in radians. The average surface brightness µ (in magnitude) of a galaxy becomes $\mu = m + 2.5\ Log(2\pi r^2)$, where *m* is the total apparent magnitude, *r* the radius. As the radius does not depend on z, from this definition it follows that the apparent surface brightness is expected to get dimmer as m, that is $\mu \sim (1+z)^{-1}$ when using standard VEGA magnitudes, or remain constant when using AB magnitudes. In the following we use AB magnitudes.

In applying this linear relation between z and d, we are not here proposing any physical model that would produce such a relation — we simply extrapolate the local properties of the Universe to see whether they are consistent with the surface brightness data.

### 3. The Samples definition.

At present, the best data set for studying the properties of objects in the distant Universe is the Hubble Ultra Deep Field (HUDF) [10], which is a set of 4 images obtained with the advanced camera for survey (ACS) in the B, V, I, and z bands down to an unprecedentedly faint apparent magnitude ($m_{AB} \sim 29$). To avoid large and uncertain k-corrections, the SB must be compared as much as possible at the same rest frame wavelengths for all objects. To satisfy this condition and properly compare galaxies up to z~5, we have chosen two reference ultraviolet bands, namely the FUV (1550 Å) and NUV (2300 Å) bands as defined by the GALEX satellite, enabling the creation of 8 pairs of samples matched to the HUDF data.

To minimize the effects of k-correction, the redshift range covered by each GALEX-HUDF pair was set requiring a maximum difference of 10% between the central rest wavelength determined by the GALEX and ACS filters. Moreover, to avoid biasing the comparison of data obtained with telescopes having different resolutions, we also require that the minimum measurable physical size of galaxies $r_m$ is the same, in each pair of samples, for GALEX (low z) and HUDF (high z). We have determined the minimum measurable angular radius of galaxies, $\theta_m$, for each of the telescopes by plotting the abundance of galaxies (with stellarity index < 0.4) vs. angular radius for all GALEX MIS3-SDSSDR5 galaxies and for all HUDF galaxies and determining the lower-cutoff angular radius for each. We took this cutoff to be the point at which the abundance per unit angular radius falls to 1/5 of the modal value. For GALEX this cutoff is at a radius of 2.4 ± 0.1 arcsec for galaxies observed in the FUV and 2.6 ± 0.2 arcsec for galaxies observed in the NUV, while for Hubble this cutoff is at a radius of 0.066 ± 0.002 arcsec, where the errors are the 1 σ statistical uncertainty. We averaged the NUV and FUV cutoffs to find the ratio of $\theta_m$GALEX/$\theta_m$HUDF to be 38$\pm$ 3. In accord with our test model, with minimum measurable physical radius $r_m \sim z\theta_m$, we chose pairs of samples so that the ratio of mean z in the HUDF sample to mean z in the GALEX sample is also as close as possible to ~38. Thus $r_m$, assuming the model, is the same for each member of the pair of samples.

In order to avoid effects due to the luminosity of galaxies, we limited objects in the samples to a narrow range of absolute magnitude M: -17.5 < M < -19.0, matching the mean absolute magnitude of each pair down to 0.02 mag, in such a way as to maximize the total number of galaxies in the pair. These are the brightest galaxies that are present in both GALEX and HUDF samples. Because galaxy size increases somewhat with absolute luminosity, these are also the galaxies most easily resolved and measured by both instruments. These UV data have the important advantage of being sensitive only to emissions from very young stars. Therefore we are in no sense looking at progenitors of GALEX galaxies, but rather at galaxies whose stellar populations are comparable in age. By analogy we are looking at populations of "babies" at different epochs in history, not comparing younger and older adults born at the same time. The important question of the comparability of the GALEX and HUDF samples is dealt with in greater detail in Section 5.3.

Finally we restricted the samples to disk galaxies with Sersic number < 2.5 so that radii could be measured accurately by measuring the slope of the exponential decline of SB within each galaxy. This measurement technique, using the slope of SB to determine radius, eliminates errors that can be introduced by measuring the radius at some arbitrarily determined isophote. For the GALEX sample, we measured radial brightness profiles and fitted them with a Sersic law, finding that nearly all these bright UV galaxies, as expected, had Sersic number < 2.5. For HUDF, we used the Sersic number provided in the HUDF catalog [11]. We also used the HUDF and GALEX catalogs to exclude all non-galaxies. The properties of the selected galaxies are summarized in Tables 1, 2 and 3.

## 4. Determination of redshift, radius and magnitude of galaxies.

For the HUDF dataset, the redshift was based on the HUDF photometric catalogs. These catalogs contain photometric measurements for each galaxy in the B, V, I, z, H and J bands. Each galaxy has a photometric redshift, estimated by two methods: Bayesian Probability (BPZ) and Maximum Likelihood (BML). Coe et al [11] report that a comparison of BPZ with spectroscopic redshifts in the small sample where they are available indicates that, except for a few outliers, BPZ redshifts are accurate to 0.04. To eliminate outliers, we have chosen to use the difference between BML and BPZ redshifts as an indicator of the reliability of BPZ redshifts, retaining only sources for which the two redshifts differ by less than 0.5. For GALEX, we limited our samples to galaxies with *spectroscopic* redshift derived from cross-correlating the MIS3 with data from the SLOAN Digital Sky Survey (SDSS) Data Release 5.

To measure total flux and half light radius, we extracted the average surface brightness profile for each galaxy from the HUDF or GALEX images. The apparent magnitude of each galaxy is determined by measuring the total flux within a fixed circular aperture large enough to accommodate the largest galaxies, but small enough to avoid contamination from other sources. To choose the best aperture over which to extract the radial profile, for each sample we compared average magnitudes and average radii as derived for a set of increasingly large apertures. We then defined the best aperture as the smallest for which average values converged. We found that these measurements are practically insensitive to the chosen aperture above this minimum value.

Finally, to determine scale-length radius, we fitted the radial brightness profile with a disk law excluding the central 0.1 arcsec for HST and 5 arcsec for GALEX , which could be affected by the PSF smearing. Given the magnitude and radius, the SB is obtained via the formulae in Section 2. A direct comparison between our measurements and those in the *i* band HUDF catalogue[11] show no significant overall differences. The SB for all selected galaxies is shown in Figure 3 plotted against redshift.

## 5. The Tolman test

*5.1 Comparison of surface brightness*

To perform the Tolman test, for each pair of data sets we compute the difference of average surface brightness between the low and high z dataset. These results are shown in Figure 4 and Table 5. The difference of SB between the pairs is always very small and no obvious trend depending on the redshift is apparent. The mean SB difference of all samples taken together, weighted by the number of galaxies in each pair, is $0.027 \pm 0.033$ mag/arcsec$^2$ ($1\sigma$ statistical uncertainty). A linear fit of SB differences with the <z> of the HUDF samples yields a slope of $\Delta$SB on <z> of $0.04 \pm 0.06$ mag/arcsec$^2$ (coefficient of correlation 0.28) and therefore is consistent with no correlation. Therefore these data are fully consistent with SB being constant in the redshift range explored.

We investigated whether the different resolutions of the two telescopes could bias the comparisons because different portions of the population distribution of SB are excluded as unresolved galaxies in the two (low-z and high-z ) samples of each pair. If, for example, in the HUDF samples most galaxies are resolved, while in the GALEX sample most galaxies are unresolved, the underlying populations of objects may have very different average surface brightness, <SB>, even if the <SB> of the resolved samples are the same. In this respect we point out that in the adopted Euclidean model, the GALEX and HUDF samples probe the same range of galaxy radius distribution.

To quantify and eliminate these possible biases we performed the Tolman test including all galaxies, both resolved and unresolved. To do this we made two justifiable assumptions. First, we assumed that the proportion of *disk* galaxies that were unresolved was the same as the proportion for *all* galaxies that were unresolved. That enabled us to estimate the number of unresolved galaxies for each sample. We computed the ratio of the number of unresolved galaxies (those with stellarity > 0.4) to the number of resolved galaxies for all galaxies within the redshift and absolute magnitude limits defined by each of the sub samples that we have selected (see Table 4). This comparison shows that there is no significant difference between the proportion of unresolved galaxies in the GALEX and HUDF datasets. Note also that, except for the HUDF FUVz sample, (where two-thirds of the galaxies are resolved), the resolved galaxies greatly outnumber the unresolved ones and 75% of all galaxies in each of the bins are resolved. This gives a preliminary indication that our analysis is not significantly affected by any biased population of galaxies due to the different resolution of the telescopes. It is also worth noting that this check is almost independent of cosmological assumptions because redshift is an observed quantity and absolute magnitude is close in the two models considered.

Second, we assumed that the SBs of all the unresolved galaxies were brighter than that of the median galaxy of the population. We then determined the median SB galaxy within each sub-sample, by ranking all measured SBs in the sample and including the estimated number of unresolved galaxies as being below (in value) the median. We then compared the median SB of the GALEX and HUDF samples within each pair as we did with the mean of the resolved galaxies. For a Gaussian distribution, or any symmetrical distribution, the mean and median values (of the whole population, resolved and unresolved) should be equal within statistical errors. The results are shown in Figure 5 and are compared with the mean SB results in Table 5. The mean of all eight differences, weighted by the number of galaxies in the pairs, is $-0.017 \pm 0.05$ mag/arcsec$^2$, the slope of $\Delta$SB on z(HUDF) is $-0.08 \pm 0.05$ mag/arcsec$^2$ with a correlation of 0.53, insignificant for 8 points even at a 5% level. This is all still completely consistent with zero difference in SB between high-z and low-z samples and with no dependence of SB on z, in accord with Tolman test predictions for a static Euclidean Universe.

We can use the median SB value to obtain an estimate of the variance within each sample by measuring the variance of all galaxies with SB more (in value) than the median SB. These variances are used to calculate the error bars (expected variance of sample median or sample mean) in Figures 4 and 5. With these variances, we can determine if the variation of the $\Delta$SB, measured either way, is greater than that expected purely from random variation in the samples. Using a chi-squared test, we see that for both methods, chi-squared is well below the 5%-probability limit of 14.1 for 7 degrees of freedom, being 9.0 using the mean SB method and 12.4 using the median SB method. Thus the null hypothesis that the differences are due only to the variability of the samples is accepted. The variances expected in sample medians are in fact somewhat underestimated since we do not take into account errors created by uncertainty in the actual number of unresolved galaxies. *We thus see that both versions of the Tolman test, either ignoring or taking into account the unresolved galaxies, are both entirely consistent with a static Euclidean Universe prediction of no variation in SB and entirely consistent with each other. Indeed, overall SB results for the GALEX and HUDF samples differ from each other by less than the statistical uncertainty of 0.03-0.05 mag/arcsec², a strikingly close agreement.*

Finally we have checked, by visual inspection of galaxies in the sample, that removing objects exhibiting signatures of interaction or merging do not change our conclusions. The selection of galaxies with disturbed morphology was performed by an external team of nine amateur astronomers evaluating the NUV images and isophote contours of all NUV-sample galaxies. Each volunteer examined the galaxies and only those considered unperturbed by more than 5 people were included in a "gold" sample. Although this procedure reduces the size of the sample, there is no significant difference of the SB-z trend.

*5.2 Is there a bias for size or surface brightness?*

We examined whether our results could be the result of an implicit selection for either surface brightness or, equivalently, for the radius of galaxies of a given intrinsic luminosity. The limited angular resolution of the observations imply that there is a minimum angular radius for resolved galaxies and thus a maximum SB for galaxies of a given M and z. As well, there is a limit on the dimmest SB that each telescope can observe, which puts a minimum on the SB that can be included in the sample. Together, these limits inevitably restrict the measured SB of any galaxy sample within a window of minimum and maximum SB. Are our results biased by these limits, simply reflecting the range of this window, and are we implicitly thus selecting for a narrow range of SB or angular size?

We can answer unequivocally that this "windowing" does *not* affect our results and that we have *not* imposed an implicit selection on radius or SB. First, we are including for the evaluation of the median SB ALL observed galaxies in the defined M and z ranges, whether or not they are resolved. Thus, as more than half of all galaxies in the sample and in each sub-sample are resolved, the value of the median SB is not affected by the maximum observable SB imposed by the telescope resolution.

Second, we note that, for the very luminous galaxies that we have chosen, the low-SB limits of both the HUDF and the GALEX MIS surveys are sufficiently far from the distribution of SB actually observed, in other words, the "window" is sufficiently wide, that these limits also have no effect on our median SBs. In figure 6 we show that for both GALEX and HUDF, the SB distribution for galaxies with -16 <M< -17 does not even begin to decrease until 28 mag/arcsec², dimmer than all galaxies in both of our samples with -17.5<M<-19 and more than 2 sigma away from the peak of the distribution for our samples.

Thus, because the bright galaxies we selected are large enough to be well resolved and bright enough not to be missed even at their largest, the measurement of the median SB is not affected by the "window" effect described above.

*5.3 Sensitivity of the results*

In implementing the Tolman test, we have taken care to match (using the SEU model) the linear resolutions, the rest wavelengths and the absolute magnitudes of the samples. How sensitive is the test to the accuracy of these matches? A comparison of FUV with NUV SB at the same z and M using the GALEX samples shows that SB in the wavelength range covered appears insensitive to $\lambda$, with a slope of only 0.35 mag/arcsec$^2$ of SB on log $\lambda$ .Thus, the 8% variance we allowed in $\lambda$ only results in SB difference of 0.01 mag, much less than statistical errors. Similarly the slope of SB on the log of the resolution in the GALEX sample is 2.2 ± 0.2 mag/arcsec$^2$, so an error of 5% in the ratio of resolutions of samples (or in determining the ratio of resolutions between HUDF and GALEX) will produce a change in SB of 0.05 mags, the same as the statistical error. Thus ratios in angular resolution in the range of 36-40 would not have a statistically significant effect. By comparison, a choice of cutoff in determining effective resolution, anywhere from ½ to 1/10 the modal value, would vary the ratios by less than ± 2%. Finally, the slope of SB on M is close to 1.0, so we did need to keep the <M> of the samples close to each other, as it would only take a change of 0.07 mag in <M> to produce the same change in <SB>. As can be seen from Table 1, the maximum difference is only 0.02 mag in <M>. Thus we conclude that our results are robust within the statistical errors.

*5.4. Effects of colors*

Since different stellar populations of low and high redshift galaxies might produce some systematic effects in the derived surface brightness, we have investigated the NUV-g colors of the selected galaxies, colors that are sensitive to the age of the stellar population.

We note that the NUV-g colors of the GALEX and HUDF samples are significantly different from each other, even if similar when compared with all galaxies, with the HUDF sample 1.3 mag bluer. However, both samples have colors typical of stellar populations with ages <1 Gyr, far separated from old, inactive galaxies. For our purposes here the key point is that for these very luminous galaxies there is no correlation at all between SB and NUV-g color, so the differences in color between HUDF and GALEX samples have no effect on our results.

To test for such correlations with the GALEX samples, we must avoid selection biases introduced by the SDSS redshift selection algorithm. The SDSS selection[12] eliminates galaxies with r-band SB > 23 mag/arcsec$^2$. This means that it also eliminates galaxies with blue NUV-g colors and relatively dim NUV SB. SDSS also eliminated galaxies with r-band radius <2 arcsecs, setting an effective SB minimum to r-band of r + 3.5 mag/arsec$^2$. This similarly eliminates galaxies with red NUV-g colors and relatively bright NUVSB. In both cases, the selection limits tend to create a spurious correlation of NUV-g and NUV-SB.

We can minimize such biases and obtain a true correlation of NUV-g and SB by limiting our test sample to the closest galaxies, with z < 0.05. These galaxies are close enough, within 200 Mpc, that none are affected by the minimum r-band radius cutoff. In addition, for near-by galaxies, the SDSS SB selection limit is somewhat relaxed. So we limit the GALEX samples to the NUVB and NUVV samples. We find that for the GALEX samples there is no correlation between SB and NUV-g color even at the 5% level. For the entire HUDF sample, we find the same lack of correlation. *Thus, differences in color between GALEX and HUDF samples have no effect on the SB comparison.* This is not particularly surprising. Since we have limited the samples in luminosity, the lack of change in SB simply means that there is no significant change in radius for these large, actively star forming galaxies with respect to the age of the stellar populations. We noted above that, for these bright galaxies, the slope of SB on M is close to 1, which means that radius does not vary greatly with absolute luminosity either, so it is not surprising that it does not vary much with color. Finally, we confirm that selection biases noted in the SDSS catalog affect only color-SB plots, not the overall <SB> of the samples, because there is no statistical difference between the comparison involving the NUVB and NUVV samples, which do not suffer from the selection, and the rest of the samples, which do.

*5.5 General remarks on analyses using size evolution*

In this paper we are examining the consistency of data on the SB of galaxies using the static Euclidean model with redshift proportional to distance. We therefore do not expect any evolutionary effects either in size or luminosity, in contrast to expectations in $\Lambda$CDM models. Not only are these in all cases galaxies whose UV radiation is dominated by young stellar populations, but in the static Euclidean model that we are testing the mean density of the universe remains a constant, so we expect no change with z among such young galaxies in size or in virial radius for a given luminosity.

The prediction of the SEU model that SB for a given absolute luminosity is constant with z is mathematically identical to the prediction that the mean physical radius R of a population of galaxies with a given absolute luminosity is also constant with z. From the assumed linear relation of redshift with distance, this model also predicts that mean angular radius for such galaxies is inversely proportional to z. So our SEU model demonstration that SB values are constant simultaneously demonstrates that mean R is also constant with z and that angular radius is inversely proportional to z, a conclusion also reached by Lopez-Correidora[13] for a lower-z sample.

In this paper, we do not compare data to the $\Lambda$CDM model. We only remark that any effort to fit such data to $\Lambda$CDM requires hypothesizing a size evolution of galaxies with z. Mathematically, in order to fit the observed constancy of SB data, any expanding universe model must require that the radii of galaxies with constant absolute luminosity evolve exactly as $(1+z)^{-1.5}$ in order to cancel out the $(1+z)^3$ SB dimming. Conversely, theories that predict some other size evolution, such as $(1+z)^{-1}$, will not fit the constant-SB data actually observed. Nor will this data be fit by any size evolution of the form $H(z)^{-a}$, where a is any constant and $H(z)$ is the Hubble parameter predicted by $\Lambda$CDM at a time corresponding to redshift z. In $\Lambda$CDM, $H(z) \sim (1+z)^{1.5}$ at high z, but diverges greatly from this value at low z. For example, size evolution proportional to $H(z)^{-1}$, advocated by Hathi et al[14], among others, predicts a difference in SB between low-z and high-z samples of ~1 mag/arcsec$^2$, at z=1, very far from the observations presented here, which show no difference in SB to within the statistical uncertainty of 0.05 mag/arcsec$^2$. We leave to further work an examination of whether a size evolution that coincidentally cancels out SB dimming is physically plausible.

# 6. Previous implementations of the Tolman test revisited

We reconsider in this section previous works where it was concluded that a static Universe is ruled out by the Tolman test. We show that, when data are consistently analyzed within the framework of the static cosmology adopted here, they agree with the expectation of no dimming. We explicitly show the details of the reanalysis for two works. For other works considering the Tolman test[14,15] similar conclusions were found.

## 6.1. Paper by Pahre, Djorgovski, and de Carvalho 1996

Pahre, Djorgovski, and de Carvalho[3] (PDdC hereafter) applied the Tolman test by studying the SB of elliptical galaxies in 3 clusters up to z=0.4. It was concluded that the data are in good agreement with the expectations for an expanding Universe, while the non-expanding model was ruled out at the better of 5-sigma significance level. We demonstrate here that this is not the case. To cope with the strong SB-radius correlation of elliptical galaxies, PDdC compared the SB at a fixed physical radius of 1 kpc computed for the expanding Universe, adopting $H_0=75$km s$^{-1}$ Mpc$^{-1}$, $\Omega_M = 0.2$, $\Omega_\Lambda = 0$. Unfortunately, they used the same SBs computed for the expanding case to test also the nonexpanding one. Clearly, to make a fair test all the transformations from apparent to physical sizes must be properly computed for the static model, again using the linear d-z relation. When this is done, we see that the SBs used by PDdC refer to physical radii of 1.4 kpc at z=0.23 and 1.7 kpc at z=0.4 (see Table 6). Here for consistency with PDdC we use $H_0=75$ km s$^{-1}$Mpc$^{-1}$. Due to this effect at z=0.4, an artificial SB dimming of ~0.5 magnitude is introduced. This is fully responsible for the failure of the non-expanding model claimed by PDdC. The corrected SBs are presented in Table 6. Note that VEGA magnitudes are used in this work, so a $(1 + z)$ dimming is expected for the static case. This is shown in Figure 7 where the corrected data for the static model are compared with the predictions. We thus conclude that when consistently analyzed these observations are in agreement with the expectation for a static Euclidean Universe.

### 6.2. Lubin and Sandage 2001

In a series of four papers with final results in Lubin and Sandage[4] (LS01 hereafter), the Tolman test was applied comparing the SB of local early type galaxies at average redshift $<z> = 0.037$, to the one of early type galaxies in three distant clusters, one at z=0.75 and two at z = 0.9. Reinforcing an initial claim presented by Sandage and Perelmuter[16], it was concluded that the $(1 + z)^{-4}$ surface brightness dimming (LS01 used standard VEGA magnitudes and $H_0 = 50$ km s$^{-1}$ Mpc$^{-1}$ and $q_0 = 1/2$) was in agreement with observations provided a significant amount of luminosity evolution was taken into account. The nonexpanding scenario was ruled out at the $10\sigma$ confidence level. These very same data are also re-discussed by Sandage[5], reaching similar conclusions.

These conclusions are not supported by the data for two main reasons. The first one is that, for the static scenario, Lubin and Sandage set the distance to d = $(c/H_0)\ln(1 + z)$, which is valid only for the Einstein-de Sitter static case. This is not the cosmology we are testing here, where the Hubble relation is hypothesized to be d= $cz/H_0$ at all redshift. The conversion factors (presented in their Table 8) to transform arc seconds to pc in the non-expanding model are therefore different in our model. The second reason is that the local sample includes only first rank cluster galaxies, while the high-z sample includes about 20 *normal* galaxies in each of three different clusters. This means that their distant galaxies are on average smaller and less luminous, and therefore are not directly comparable to local ones because of the well known absolute magnitude-SB relation.

LS01 presented magnitudes and SB as derived for four different Petrosian radii, as defined by the $\eta$ parameter, because a dependence on dimming with $\eta$ was found. (By definition $\eta$ is the difference in magnitudes between the surface brightness averaged over a radius, to the surface brightness at that radius. Larger $\eta$ corresponds to larger radii. For reference, the commonly used half-light radius corresponds to $\eta$ = 1.4. ) Contrary to LS01, in the static cosmology we are using, there is no difference in considering different values of $\eta$; thus in the following we limit ourselves to $\eta = 2$, which was indicated by LS01 as the most appropriate for implementing the Tolman test. The distribution of absolute magnitudes derived in our static cosmological model for the local and distant samples is shown in Figure 8. As expected there is a clear offset in luminosity between samples, local galaxies being on average 1.5 magnitudes brighter then the distant ones. Thus, to cope with the strong SB-luminosity relation we are forced to cut samples, considering only the region of overlap in luminosity, namely $-23.8 < M < -22.7$. We stress that the luminosity offset is the same on both the expanding and static scenario because, as pointed out in section 2, the luminosity distance is virtually the same in the two models. Thus the limitation in luminosity range is legitimate and is not biasing the test.

In Fig. 9 we plot the SB of the 14 selected galaxies as a function of their size (computed in the static scenario). Using Vega mag a dimming of a factor (1+z) is expected, thus data have been made brighter by this amount to be directly comparable. Within the intrinsic spread of the SB-size relation, the match between local and distance samples is good. There is only one clear outlier, the galaxy with the brightest surface brightness, which is the first entry in Table 2 of LS01. Excluding this outlier, we find a probability of 68% for the samples to be drawn from the same population. This was computed assuming a constant uncertainty of 0.15 mags on the SB, certainly an underestimation of the true uncertainty considering the large number of transformations required to convert observed quantities to the same rest frame system. Including the outlier galaxy decreases the probability but the samples remain statistically indistinguishable.

The samples overlap neatly in size, which is to say (in the non-expanding scenario) that galaxies of similar luminosity also have similar physical size and therefore, necessarily, the same SB. This is not the case in the expanding scenario, where samples do not overlap at all in radius, forcing Lubin and Sandage to extrapolate the local sample to small radii using data from Sandage and Perelmuter [16]. We conclude that far from disproving a non-expanding cosmology, data by Lubin and Sandage agree very well with predictions for a static Euclidean universe. This result effectively extends our own results, as discussed above, to early-type galaxies.

## 7. Conclusions

We find that the UV surface brightness of luminous disk galaxies are constant over a very wide redshift range (from z = 0.03 to z ~ 5). From this analysis we conclude that the Tolman test for surface brightness dimming is consistent with a non-expanding, Euclidean Universe with distance proportional to redshift. This result is also consistent with previously published datasets that were obtained to perform the Tolman test for a smaller redshift baseline  when analysis of such data is done in a consistent system.

We stress that our analysis compared samples of galaxies that were matched for:

- mean absolute magnitude,
- rest-frame wavelength,
- minimum measurable physical radius,

thus removing the needs for complex and uncertain corrections. There is no implicit or explicit bias for SB or galaxy radius.

We also emphasize that this matching of observations and predictions of the non-expanding, Euclidean Universe involves neither fitting of parameters nor any free variables. The simple prediction of constant SB, and equivalently, no size evolution in these young galaxies is consistent with all observations.

We have confirmed the constancy of SB using two statistical methods for determining mean SB of a population, one of these methods including unresolved galaxies. A re-analysis of earlier data for elliptical galaxies, covering a different range of redshift, obtained with different methods, and in different wavelengths, shows consistency with our results, thus extending the significance of the test.

The agreement of the SB data with the hypotheses of a non-expanding, Euclidean Universe and of redshift proportional to distance is not sufficient by itself to confirm what would be a radical transformation in our understanding of both the structure and evolution of the cosmos and of the propagation of  light. However, this consistency is more than sufficient reason to examine further this combination of hypotheses.

*Acknowledgments.* The authors thank Alex Jarvie and Reece Arnott for their help with software and Ralph Biggins, Aaron Blake, Allan Brewer, Harry Costas, Keith Hitchman, Hermann Hubschle, Gregory Landis, Stefan Meyer and Felix Scholkmann for help in the gold sample selection process. We also thank Dr. Timothy E. Eastman for his careful review and helpful suggestions on the manuscript.

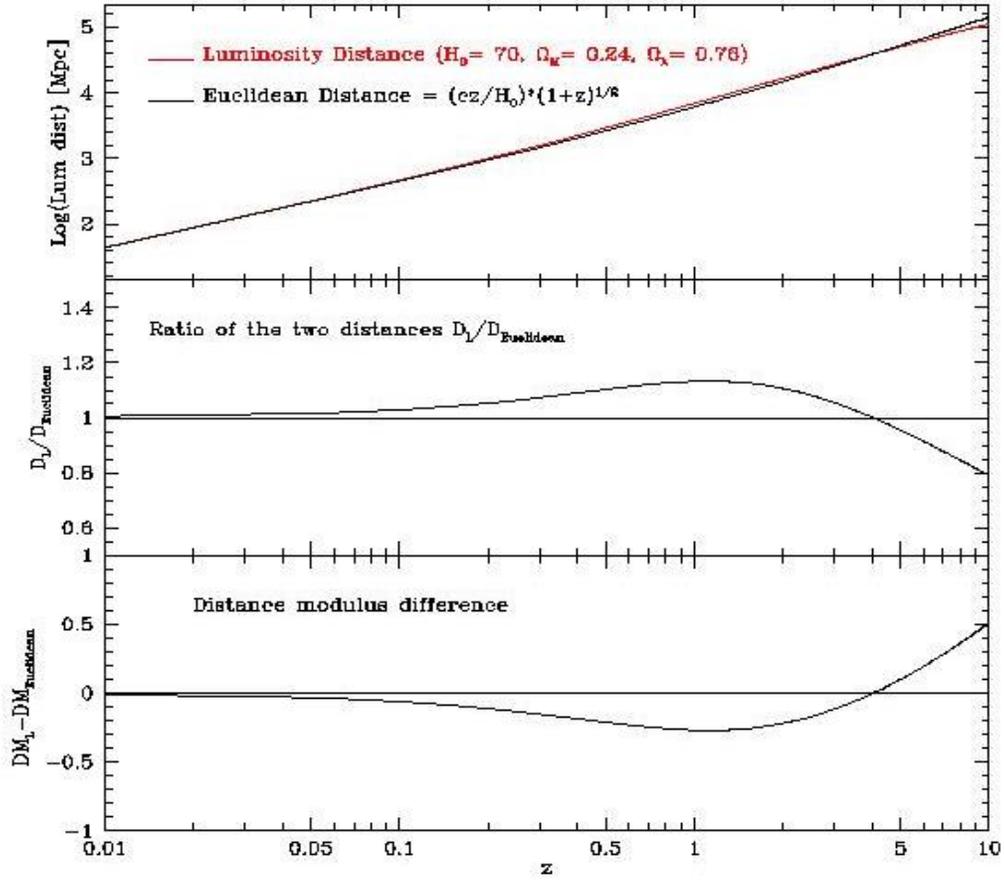

**Figure 1**

Comparison of the distance modulus for Vega magnitudes for the adopted Euclidean non-expanding universe with linear Hubble relation cosmology and the concordance cosmology. Upper panel: The distance modulus $(m - M) = 25 + 5\mathrm{Log}(cz/Ho) + 2.5\mathrm{Log}(1+z)$, where $H_0 = 70$ in km s$-1$ Mpc$-1$ as a function of the redshift z for an Euclidean Universe with $d = cz/H_0$ (black line) compared to the one obtained from the concordance cosmology with $\Omega_m = 0.26$ and $\Omega_\Lambda = 0.76$ (red line). Middle panel: Ratio of the two distances(concordance/Euclidean). Lower panel: Distance modulus difference in magnitudes(concordance-Euclidean). This graph shows clearly the similarity of the two, making galaxy selection in luminosity model-independent.

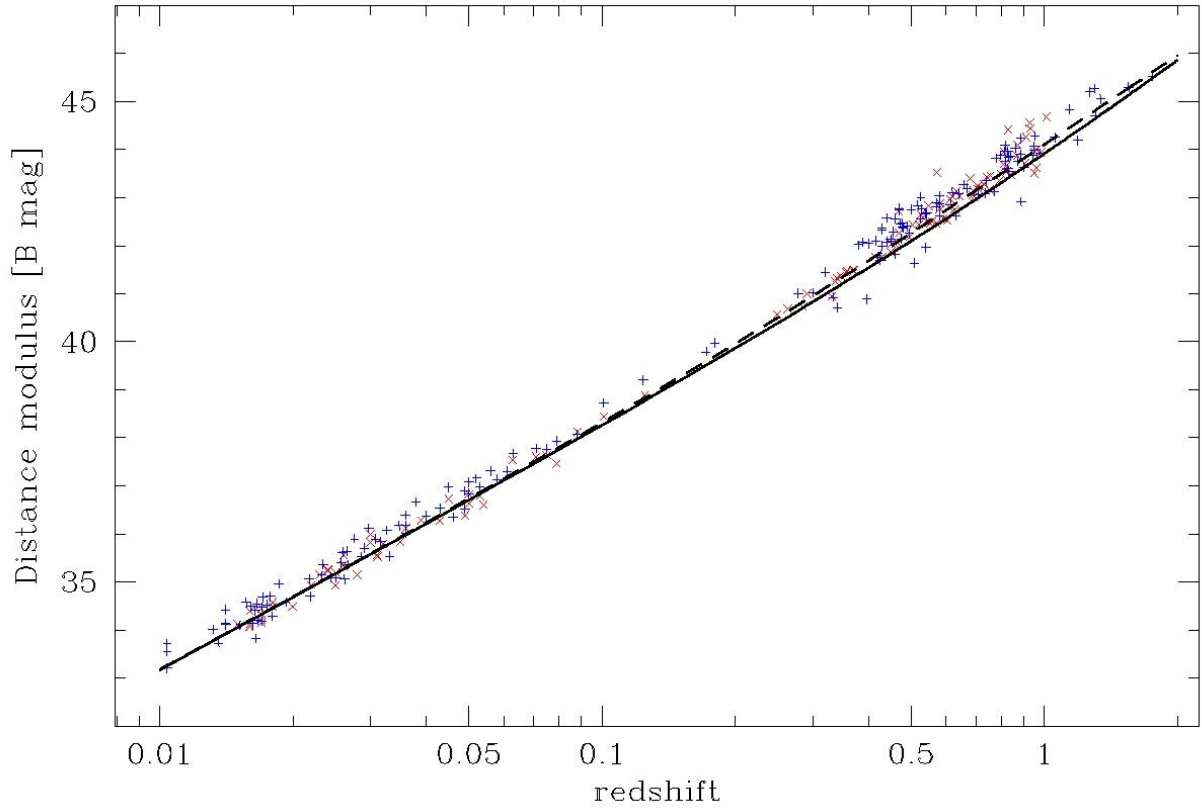

**Figure 2**

Superposed to the models are data for supernovae type Ia from the gold sample as defined in Riess et al. [8] (pluses), and the supernovae legacy survey[9], (crosses). The assumed absolute magnitude of the supernovae is M= -19.25. The two lines (SEU model with solid line and $\Lambda$CDM concordance cosmology as dashed line) are nearly identical over the entire redshift range, differing at no point by more than 0.15 mag and in most of the region by less than 0.05 mag.

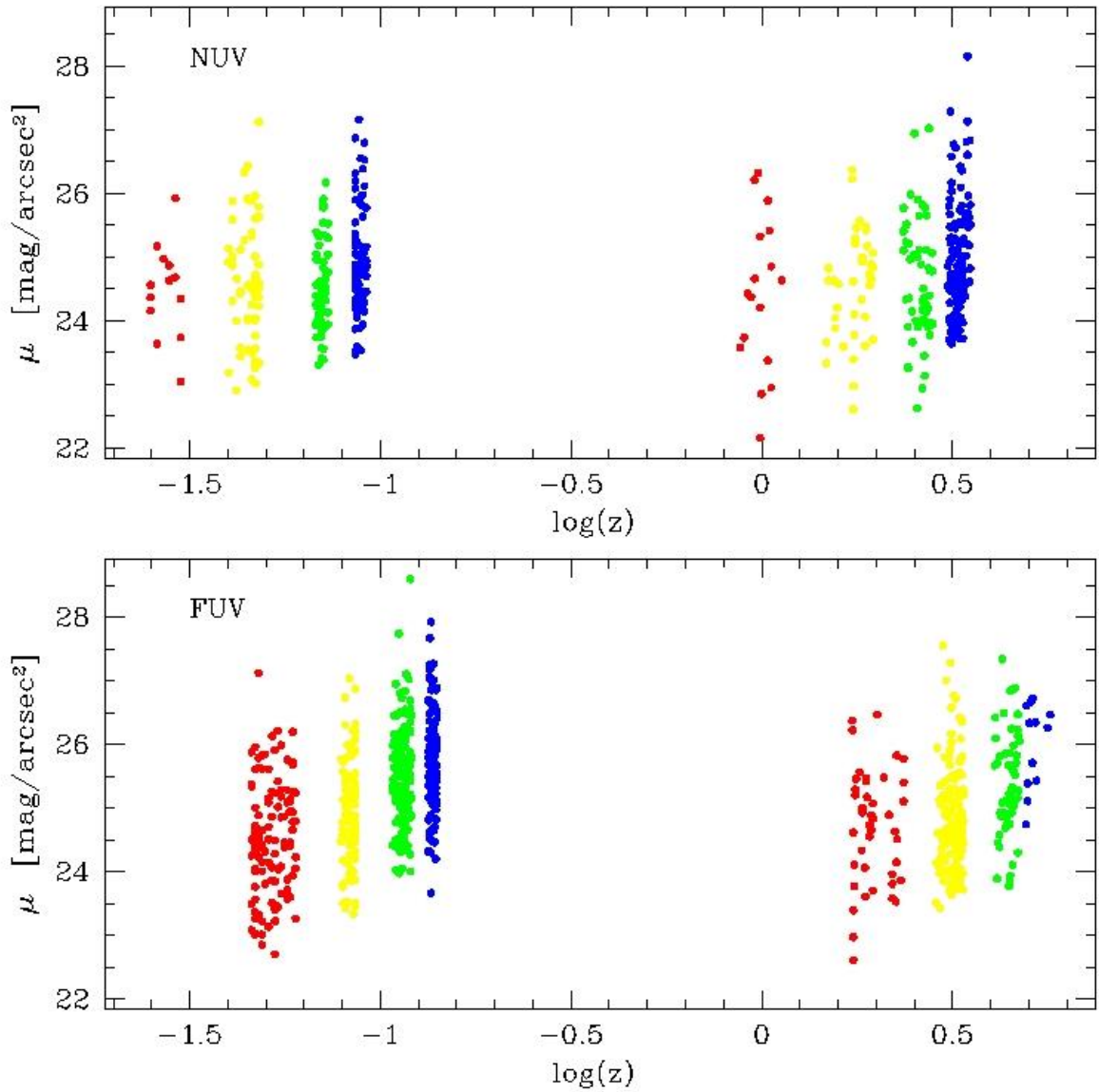

**Figure 3**. The average surface brightness of individual galaxies for the local GALEX and high z HUDF samples as a function of redshift (Near UV upper panel; Far UV lower panel). The trend of increasing SB with z within the high and low-z datasets is due to the effect of limited telescope resolution.

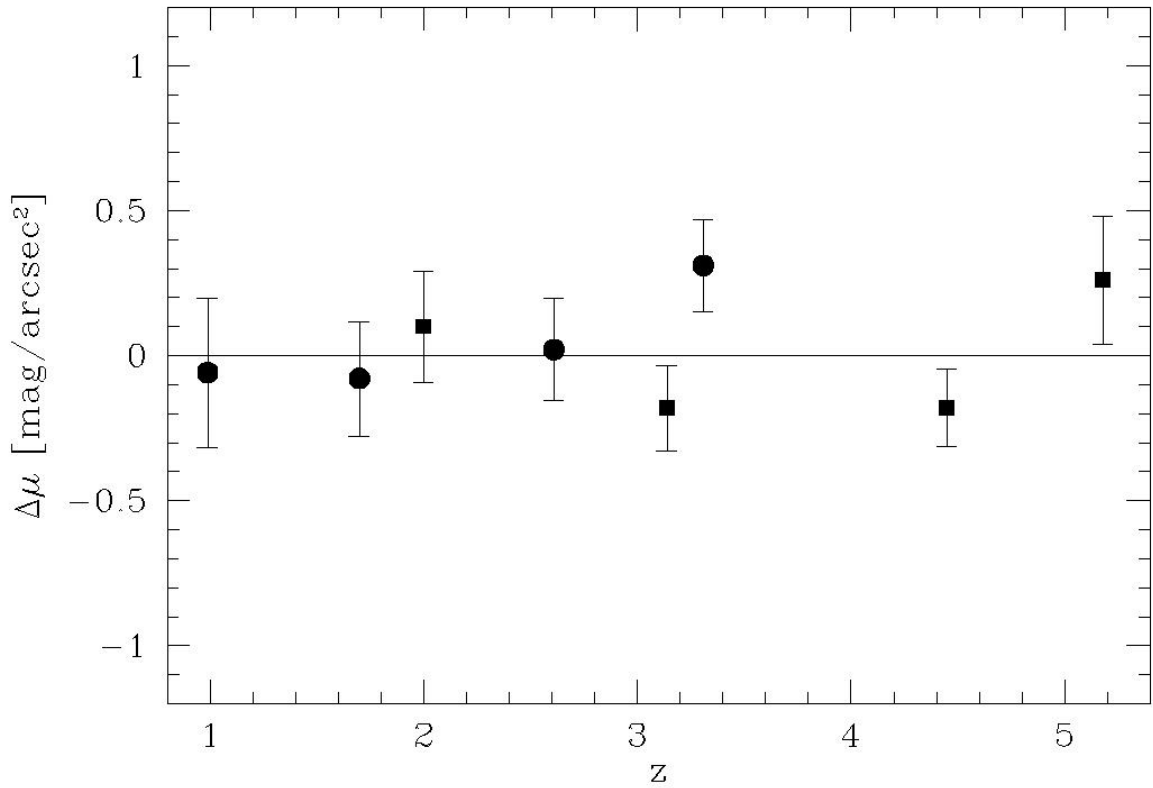

**Figure 4.** The difference in mean SB ($\Delta\mu = \mu_{HUDF} - \mu_{GALEX}$) between the HUDF and GALEX members of each pair of matched samples plotted against the mean redshift of the HUDF samples (filled circles NUV dataset, filled squares: FUV dataset). Results are consistent with no change in SB with z. Error bars are 1-sigma statistical errors.

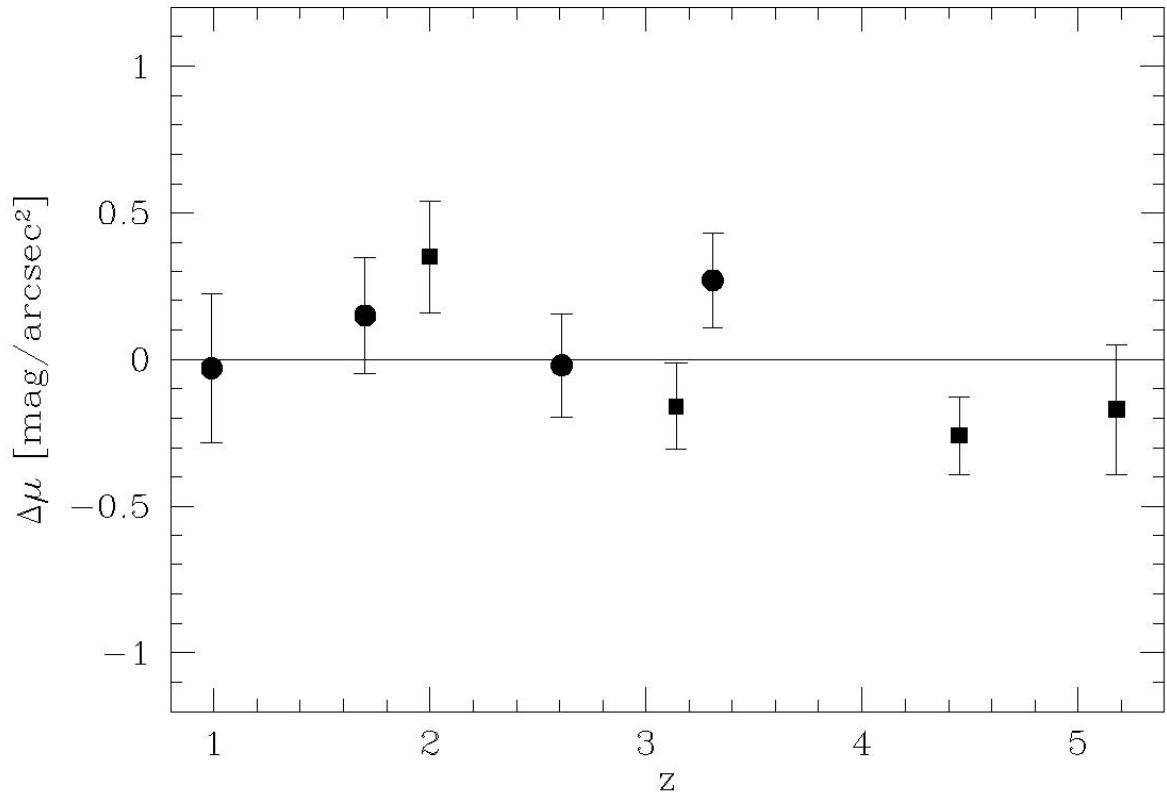

**Figure 5.** The difference in median SB (taking into account unresolved galaxies) between the HUDF and GALEX members of each pair of matched samples is plotted against the mean z of the HUDF sample (filled circles NUV dataset, filled squares: FUV dataset). As with the mean SB, results are consistent with no change in SB with z. Error bars are one-sigma statistical errors.

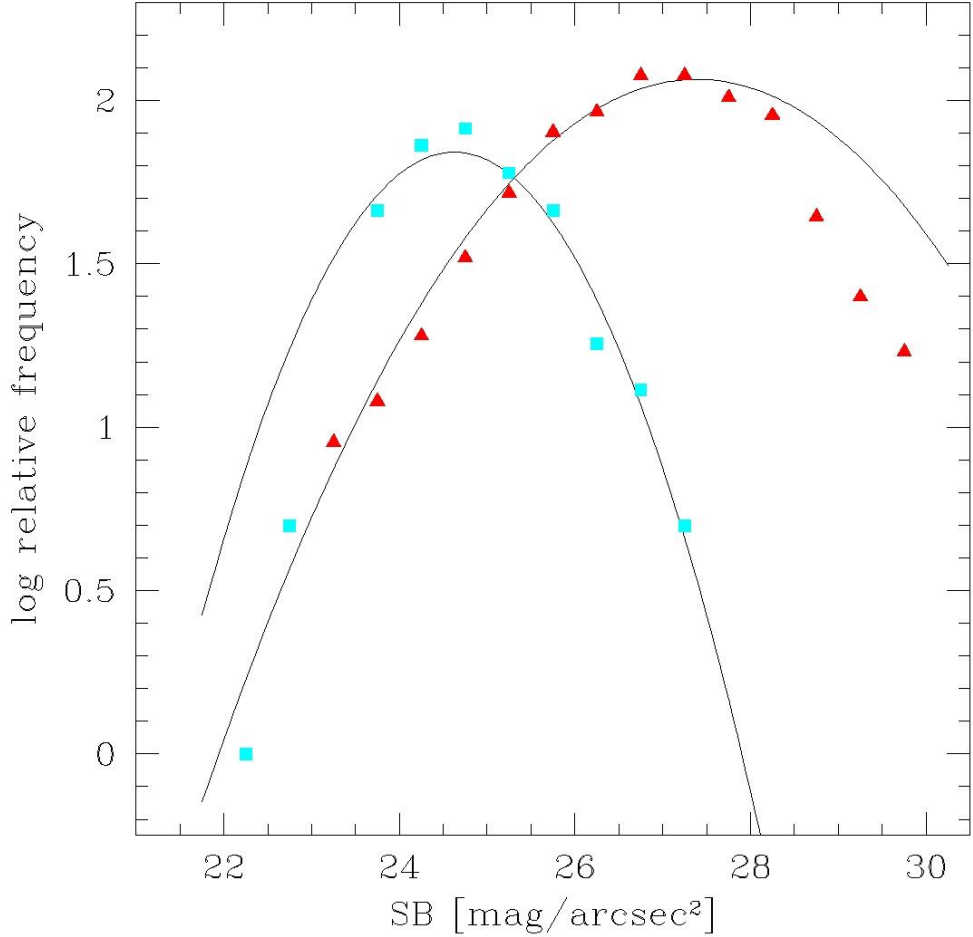

**Figure 6.** Log relative frequency of galaxies are plotted against SB for the selected HUDF sample with -17.5<M<-19 (squares) and with -16<M-17 (triangles). The dimmer galaxies on the right show the effect of the limits of SB visibility is significant only for galaxies dimmer than 28.5 mag/arcsec² which does not affect the distribution of the galaxies in the sample. The sample galaxies on the left show a similar cutoff due to the smallest, highest SB galaxies being unresolved. In both cases the curves are Gaussian fits to the non-cutoff sections of the distributions.

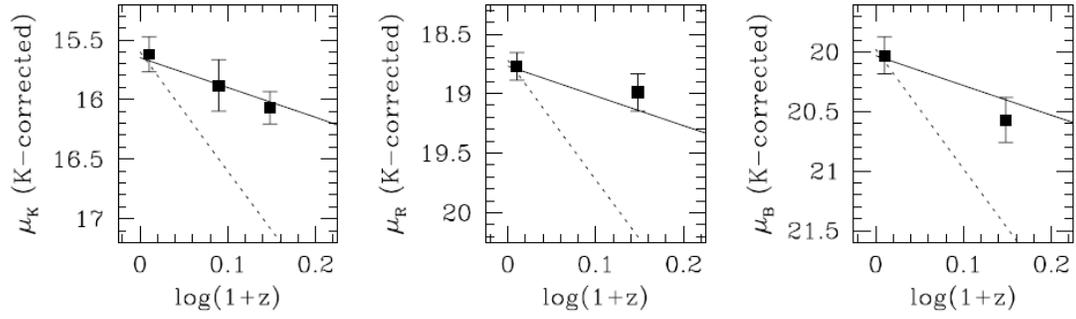

**Figure 7**. Tolman SB test for the non-expanding Universe scenario in K, R, and B bands as derived from PDdC[3] data. Values now correctly refer to the same physical radius of 1 kpc (in the non-expanding scenario). Data have been K-corrected (with the same values used in PDdC); thus SBs are expected to follow and do follow the $(1+z)^{-1}$ trend (solid line). The dotted lines show the $(1+z)^{-4}$ dimming expected in the expanding Universe scenario.

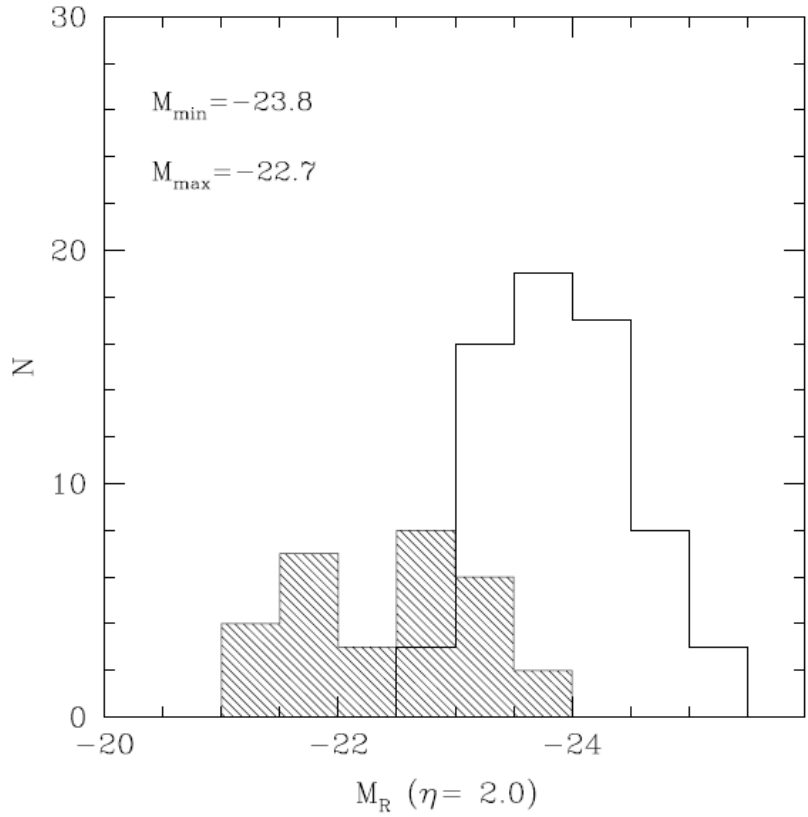

**Figure 8**. Distribution of R-band absolute magnitudes for galaxies studied by Lubin and Sandage [4] for $\eta =$ 2 (see text). Local (empty histogram) and distant (shaded histogram) galaxies have substantially different luminosities. The given maximum and minimum absolute magnitudes define the region of overlap in luminosity of the two samples used to select the galaxies shown in Figure 9.

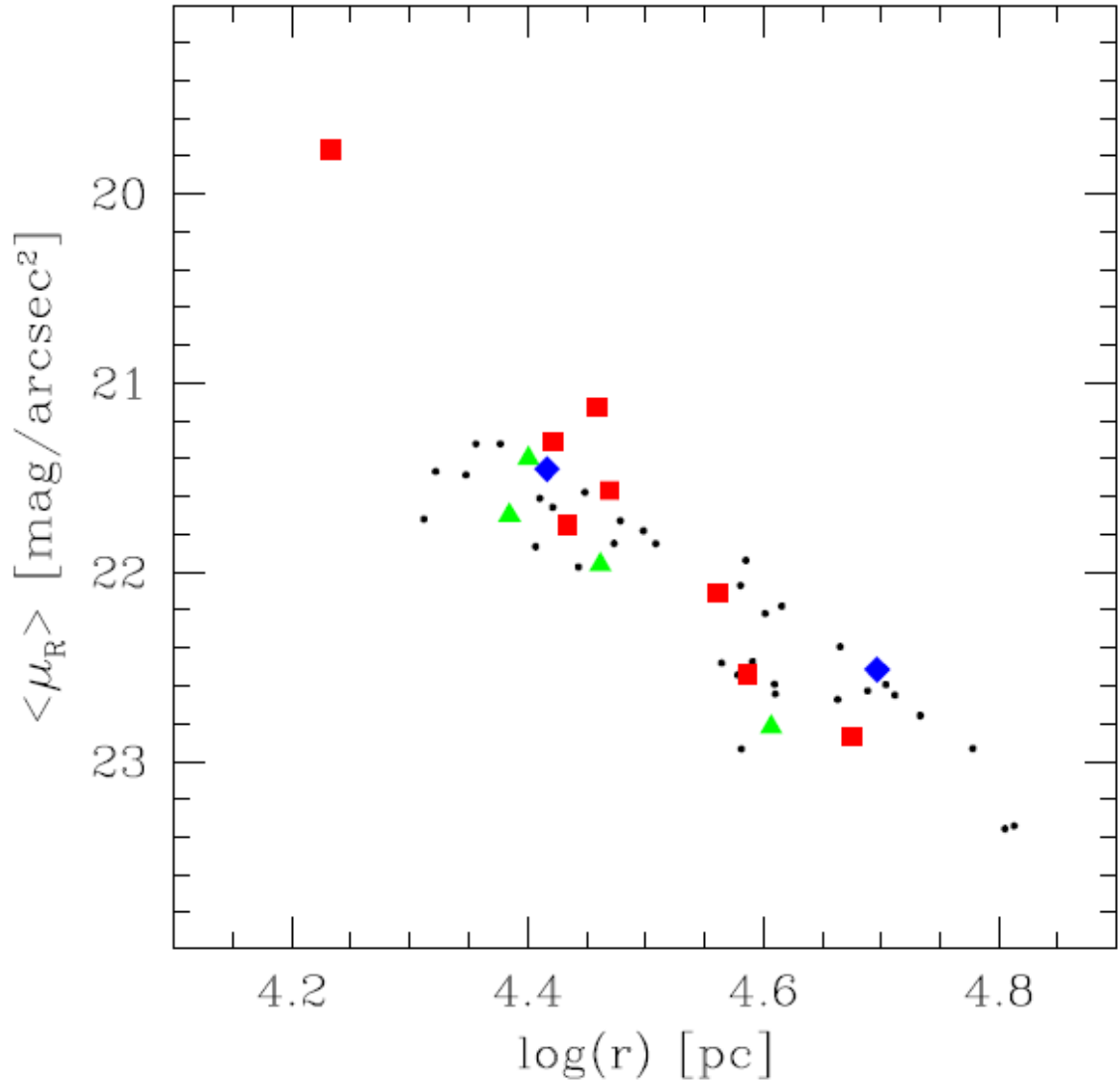

**Figure 9.** Tolman SB test for a non-expanding Euclidean Universe as derived from Lubin and Sandage [4] data, restricted to the region of overlap in luminosity between local and distant samples. Surface brightness in VEGA mag/arcsec² is k-corrected and made brighter by a factor 2.5Log(1+z) to remove the expected dimming. Thus, in absence of expansion, galaxies of the same luminosity should have the same

SB within the scatter of the SB-luminosity relation. They should also have comparable physical radius. The transformation from arcsec to pc was done assuming $d = cz/H_0$ for all redshifts. Different symbols refers to the local sample at $z = 0.037$ (Dots), galaxies in cluster Cl 1604+43 at $z = 0.924$ (squares), cluster Cl 1604+43 at $z = 0.897$ (Diamonds), and cluster Cl 1324+30 $z = 0.756$ (Triangles).

| HUDF | | | | | | |
|---|---|---|---|---|---|---|
| Sample | <z> | $Z_{min}$ | $Z_{max}$ | <λ> | <M> | N |
| | | | | | | |
| NUVB | 0.99 | 0.88 | 1.13 | 217 | -18.08 | 17 |
| NUVV | 1.70 | 1.48 | 1.96 | 219 | -18.12 | 23 |
| NUVi | 2.61 | 2.36 | 2.81 | 213 | -18.25 | 53 |
| NUVz | 3.31 | 3.08 | 3.54 | 210 | -18.23 | 73 |
| | | | | | | |
| FUVB | 2.00 | 1.73 | 2.36 | 144 | -18.10 | 28 |
| FUVV | 3.14 | 2.87 | 3.38 | 143 | -18.27 | 86 |
| FUVi | 4.45 | 4.09 | 4.75 | 141 | -18.29 | 52 |
| FUVz | 5.18 | 4.94 | 5.72 | 146 | -18.22 | 12 |
| GALEX | | | | | | |
| | | | | | | |
| NUVB | 0.027 | 0.025 | 0.030 | 224 | -18.06 | 13 |
| NUVV | 0.045 | 0.040 | 0.048 | 220 | -18.12 | 43 |
| NUVi | 0.070 | 0.068 | 0.073 | 215 | -18.27 | 64 |
| NUVz | 0.089 | 0.086 | 0.092 | 211 | -18.23 | 80 |
| | | | | | | |
| FUVB | 0.053 | 0.046 | 0.060 | 147 | -18.11 | 87 |
| FUVV | 0.083 | 0.079 | 0.086 | 143 | -18.27 | 73 |
| FUVi | 0.114 | 0.108 | 0.121 | 139 | -18.28 | 148 |
| FUVz | 0.137 | 0.134 | 0.141 | 136 | -18.21 | 115 |
| | | | | | | |
| | | | | | | |

**Table 1** Col. 1 is the sample band designation, with HUDF samples first and GALEX last. Col. 2 is the mean redshift of the sample; Col. 3 is the minimum z; Col. 4 is the maximum z; Col. 5 is the center of the wavelength band observed at the galaxy (rest frame); Col. 6 is the mean absolute magnitude of galaxies in the sample, and Col. 7 is the number of galaxies in the sample.

| Sample | Total in z-M range | Galaxies with Stellarity>0.4 | Sersic index < 2.5 | Galaxies Used |
|--------|--------------------|-----------------------------|--------------------|---------------|
| NUVB | 27 | 25 | 25 | 13 |
| NUVV | 80 | 74 | 73 | 43 |
| NUVi | 141 | 117 | 117 | 64 |
| NUVz | 154 | 119 | 119 | 80 |
| FUVB | 137 | 128 | 126 | 87 |
| FUVV | 87 | 77 | 77 | 73 |
| FUVi | 189 | 156 | 156 | 148 |
| FUVz | 157 | 132 | 130 | 115 |

**Table 2.** GALEX selection table — Col. 1 is the sample name; Col. 2 is the total number of GALEX MIS3 galaxies in the z-M window for the sample; Col. 3 is the number of galaxies resolved (stellarity>0.4). Col 4 is the number of galaxies with Sersic index < 2.5; Col. 5 is the number in the samples actually compared, after eliminating galaxies with neighbors within 10 pixels, with artifacts in the same radius, and by eliminating lowest or highest absolute magnitude galaxies in a sample to achieve identical <M> in the GALEX and HUDF members of each sample pair. A large fraction of lower-brightness GALEX galaxies had to be eliminated in each case because the abundance of GALEX galaxies brighter than -18 falls off much faster with increasing brightness than for the HUDF sample.

| Sample | Total in z-M | Galaxies good z | Stellarity <0.4 | ΔSersic Index <1.0 | Sersic Index <2.5 | Galaxies unresolved | Galaxies used |
|--------|------|------|------|------|------|------|------|
| | | | | | | | |
| NUVB | 32 | 32 | 31 | 28 | 21 | 4 | 17 |
| NUVV | 66 | 62 | 61 | 46 | 29 | 12 | 23 |
| NUVi | 103 | 88 | 84 | 64 | 58 | 10 | 53 |
| NUVz | 180 | 175 | 154 | 118 | 96 | 20 | 72 |
| | | | | | | | |
| FUVB | 62 | 54 | 53 | 43 | 41 | 7 | 28 |
| FUVV | 169 | 168 | 157 | 116 | 107 | 18 | 86 |
| FUVi | 148 | 148 | 119 | 73 | 68 | 14 | 52 |
| FUVz | 74 | 62 | 42 | 13 | 13 | 19 | 12 |

**Table 3.** HUDF selection table — Col. 1 is the sample name; Col. 2 is the total number of HUDF galaxies in the z-M window for the sample; Col. 3 is the number of galaxies with good z (see text); Col. 4 is the number of galaxies with stellarity<0.4; Col. 5 is the number of galaxies with uncertainty in Sersic number < 1.0; Col. 6 is the number of galaxies with Sersic < 2.5 (disk galaxies); Col. 7 number of unresolved galaxies; Col. 8 is the number of galaxies in the samples actually compared, after eliminating galaxies with no Sersic determinations, with neighbors within 10 pixels, and by eliminating lowest or highest absolute magnitude galaxies to achieve identical <M> in the GALEX and HUDF sample pairs.

| Sample | HUDF unresolved ratio | Galex unresolved ratio |
|---|---|---|
| NUVB | 0.03 | 0.08 |
| NUVV | 0.01 | 0.08 |
| NUVi | 0.08 | 0.21 |
| NUVz | 0.14 | 0.29 |
| | | |
| FUVB | 0.04 | 0.07 |
| FUVV | 0.07 | 0.13 |
| FUVi | 0.25 | 0.21 |
| FUVz | 0.50 | 0.19 |

**Table 4.** Col. 1 is the name of the HUDF-GALEX comparison pair; Col.2 and 3 is the ratio of unresolved (stellarity > 0.4) HUDF (Col 2) and GALEX (Col 3) galaxies to resolved galaxies in a redshift-absolute magnitude bin equal to that of the member sample of each pair. The ratios are similar in both columns for each of the eight pairs, except for the most distant one, FUVz, indicating that a similar part of the SB distribution is sampled in the HUDF and GALEX samples. Also, again excepting the HUDF FUVz sample, the numbers are all <0.30, showing that >75% of all galaxies in the bins are resolved.

| Sample | Mean SB | | | Median SB | | |
|---|---|---|---|---|---|---|

| | HUDF ⟨μ⟩ | GALEX ⟨μ⟩ | Δ⟨μ⟩ | $\chi_v^2$ (mean) | HUDF Med μ | GALEX Med μ | Δ Med SB | $\chi_v^2$ (med) |
|---|---|---|---|---|---|---|---|---|
| NUVB | 24.41 | 24.47 | -0.06 | 0.06 | 24.43 | 24.46 | -0.03 | 0.02 |
| NUVV | 24.45 | 24.53 | -0.08 | 0.16 | 24.58 | 24.43 | 0.15 | 0.57 |
| NUVi | 24.56 | 24.54 | 0.02 | 0.01 | 24.30 | 24.32 | -0.02 | 0.01 |
| NUVz | 25.19 | 24.88 | 0.31 | 3.73 | 24.85 | 24.58 | 0.27 | 2.83 |
| | | | | | | | | |
| FUVB | 24.70 | 24.60 | 0.10 | 0.27 | 24.85 | 24.50 | 0.35 | 3.35 |
| FUVV | 24.75 | 24.93 | -0.18 | 1.62 | 24.61 | 24.77 | -0.16 | 1.28 |
| FUVi | 25.34 | 25.52 | -0.18 | 1.81 | 25.06 | 25.32 | -0.26 | 3.77 |
| FUVz | 25.98 | 25.72 | 0.26 | 1.38 | 25.38 | 25.55 | -0.17 | 0.59 |
| **Total** | | | | **9.04** | | | | **12.42** |

**Table 5.** Col. 1 is sample pair name; Col. 2 and Col. 3 are the mean SB for the HUDF and GALEX members of the pair respectively; Col. 4 is the difference (HUDF-GALEX) in mean SB and Col. 5 is the chi square for this difference. Col. 6 and Col. 7 are the median SB for the HUDF and GALEX members of the pair respectively; Col. 8 is the difference (HUDF-GALEX) in median SB and Col. 9 is the chi square for this difference. The $\chi_v^2$ totals shown are consistent with the difference in both mean and median SB being entirely due to the statistical variance of the samples.

| Cluster | Z | Band | <μ>e | Radius | Slope | μ | <μ>e | K-corr | <μ>e |
|---------|---|------|------|--------|-------|---|------|--------|------|
|  |  |  |  | (kpc) |  | (1 kpc) |  |  |  |
| *1* | *2* | *3* | *4* | *5* | *6* | *7* | *8* | *9* | *10* |
| Coma | 0.024 | K | 15.63 ± 0.11 | 1.04 | −1.0 | −0.04 | 15.59 | −0.03 | 15.62 ± 0.15 |
| Abell 2390 | 0,159 | K | 16.01 ± 0.19 | 1.37 | −1.0 | −0.34 | 15.67 | −0.21 | 15.88 ± 0.21 |
| Abell 851 | 0,282 | K | 16.36 ± 0.09 | 0.09 | −1.0 | −0.56 | 15.80 | −0.27 | 16.07 ± 0.14 |
|  |  |  |  |  |  |  |  |  |  |
| Coma | 0.024 | R | 18.84 ± 0.06 | 1.04 | −0.9 | −0.05 | 18.79 | +0.02 | 18.77 ± 0.12 |
| Abell 851 | 0,282 | R | 20.30 ± 0.12 | 1.68 | −0.9 | −0.63 | 19.67 | +0.62 | 19.05 ± 0.16 |
|  |  |  |  |  |  |  |  |  |  |
| Coma | 0.024 | B | 20.19 ± 0.12 | 1.04 | −0.9 | −0.05 | 20.14 | +0.11 | 20.03 ± 0.16 |
| Abell 851 | 0,282 | B | 22.95 ± 0.16 | 1.68 | −0.9 | −0.63 | 22.32 | +1.75 | 20.57 ± 0.19 |

**Table 6.** Column 1, sample name; Column 2, redshift; Column 3, spectral band observed; Column 4, the observed surface brightness in mag/arcsec2, as reported by PDdC. This value is not K-corrected. Column 5, the radius at which the SB quoted by PDdC refers in the non-expanding Universe. Column 6, the slope of the relation between radius and surface brightness in flux units $r_e \propto \Sigma^a$ as quoted in PDdC. Column 7, the correction in magnitudes to bring the SB in Column 4 back to 1 kpc. Column 8, the corrected value for the surface brightness at 1 kpc in the non-expanding Universe; Column 9, the k-correction as in PDdC; Column 10, the final value for the surface brightness at 1 kpc including both Δμ and k-correction. Errors include in quadrature a 0.1 mag to accommodate uncertainties on the k-correction and the slope of the $r_e \propto \Sigma$ relation; d=cz/H0 is applied at all z.